\documentstyle[aps,multicol,epsfig,prb]{revtex}

\input epsf

\def\prb{Phys.\ Rev.\ {\bf B}}
\def\prl{Phys.\ Rev.\ Lett.\/}

\def\Cal{\cal}

\newcommand{\be}{\begin{equation}}
\newcommand{\ee}{\end{equation}}

\newcommand{\e}[1]{\epsilon_{\vec{k}}}

\newcommand{\ba}{\begin{eqnarray}}
\newcommand{\ea}{\end{eqnarray}}

\begin{document}

\title
{Quantum Theory of a Nematic Fermi Fluid. }
\author{Vadim Oganesyan$^{1}$, Steven A. Kivelson$^{1}$, Eduardo Fradkin$^{2}$}
\address
{Dept.\ of Physics, U.\ C.\ L.\ A.$^{1}$, Los Angeles, CA  90095;  
 Dept.\ of Physics, University of Illinois$^{2}$, Urbana, IL 61801}
\date{\today}
\maketitle 

\begin{abstract}

We develop a microscopic theory of the electronic nematic phase 
proximate to an  isotropic Fermi liquid in both two and three
dimensions. 
Explicit expressions are obtained for the small amplitude collective excitations
in the ordered state; remarkably, the
nematic Goldstone mode (the directorwave) is overdamped except along special 
directions dictated by symmetry.  At the quantum critical point we find a 
dynamical
exponent of $z=3$, implying stability of the gaussian fixed point.
The leading perturbative effect of the overdamped Goldstone modes leads
to a breakdown of Fermi liquid theory in the nematic phase 
and to strongly angle dependent electronic self energies around the Fermi 
surface.
Other metallic liquid crystal phases, {\it e.\ g.\/} a quantum hexatic, behave 
analogously.

\end{abstract}


\begin{multicols}{2}
\narrowtext

There is a growing body of both experimental 
and theoretical evidence for the relevance of inhomogeneous and/or anisotropic 
metallic phases in a wide array of highly correlated electronic systems.  
Quasi-one dimensional (stripe or ``electronic smectic'') phases have been 
observed in a large variety of transition metal oxides.\cite{stripes}  
More recently,\cite{lilly,stormer} the dramatic discovery of a metallic phase 
with
a strongly anisotropic resistivity tensor for a range of magnetic fields
in ultra clean heterojunctions has provided clear evidence of the
existence of a ``quantum Hall nematic'' phase.
In parallel, theoretical work\cite{KEF,FK} has been carried out on electronic 
liquid crystal phases.
These ground-state phases are classified, based on broken symmetries,
by analogy with  classical liquid crystals.  So far, these studies have
focused primarily on the
smectic, which is a unidirectional density wave 
with broken translational symmetry in only one direction, but which supports 
liquid-like
electron flow\cite{EFKL,qhsmectic}, and to a lesser extent on the 
nematic, which is uniform but anisotropic (breaks 
rotational symmetry)\cite{FKMN,WD}.
A nematic state in the 
proximity to the smectic state can be visualized most naturally as a 
melted smectic, {\it i.\ e.\/} a smectic with dislocations.  However, 
a theory of the nematic phase based on this picture has yet to be satisfactorily
formalized.

In this paper we approach the nematic metal via a complimentary route, from the 
isotropic and weakly correlated side. In this limit, the zero temperature 
isotropic 
to nematic transition is a Fermi surface instability.   
The director order parameter which characterizes the broken symmetry of the 
nematic state
is a rank two  symmetric traceless tensor which is even under time reversal.  In 
two
dimensions, but not in three, the order parameter is odd under 90$^\circ$ 
spatial rotation.
For simplicity we first consider  spinless fermions 
in two dimensions with full rotational symmetry, deferring until later any 
discussion of spin and the
symmetry breaking effects of the crystal fields which are inevitable in actual 
solids. While
many microscopic definitions of this order parameter are possible, we shall see 
that the
natural one in the present context is the quadrupole  density
\ba
\hat{\bf Q}(x)&\equiv&
-\frac{1}{k^2_F} {\Psi}^{\dagger}(\vec{r})
\left( 
\begin{array}{cc}
\partial^2_x-\partial^2_y & 2\partial_x\partial_y\\ 
2\partial_x\partial_y     & \partial^2_y-\partial^2_x
\end{array}
\right) \Psi(\vec{r}),
\label{Q}
\ea
where $k_F$ is the Fermi wave number.
The order parameter 
${\bf Q}\equiv <\hat{\bf Q}>$ can be expressed in terms of an 
amplitude and a phase,
$Qe^{2i\theta}=Q_{11} + i Q_{12}$.  Thus, in the broken symmetry state   
the Fermi surface is elliptical ({\it i.e.} the Fermi momentum varies around the 
Fermi surface),
with eccentricity  proportional to $Q$ along a major axis
at an azimuthal angle $\pm \theta$.

At finite temperature, $T$, where the long-distance physics is purely classical,
an electronic liquid crystal has much the same character as a conventional 
liquid crystal.
In the case of the two dimensional nematic, this means that there is no true 
long-range order,
and a Kosterlitz-Thouless transition at a critical temperature to the disordered 
(high
temperature) state.  It is a remarkable feature of the hydrodynamics of 
classical nematics (in
both two and three dimensions) that the director wave ({\it i.e.} the finite 
temperature
 Goldstone mode) is overdamped\cite{forster,CL}.

At zero temperature, the nematic phase possesses a true broken symmetry.  
Many features of this state follow as a direct consequence of symmetry breaking, 
independent of microscopic
considerations.  The quantum transition between the nematic and isotropic states 
can be studied
at mean field level by considering the Landau expansion of the ground-state 
energy as
\be
E({\bf Q})=E({\bf 0})+\frac{A}{4}{\rm Tr}[{\bf Q}^2]+ 
\frac{B}{8}{\rm Tr}[{\bf Q}^4]+\ldots
\label{GS2D}
\ee
 That only even terms appear in
this expansion is a consequence of the odd parity of ${\bf Q}$ under 90$^\circ$ 
rotations.  
At the transition point, $A$ changes
from positive in the isotropic phase to negative in the nematic phase, in which
$Q=\sqrt{|A|/B} + \ldots$.  The elastic theory which governs any long wavelength, static
variations of the order parameter is directly inherited from the classical 
theory\cite{forster,CL},
leading to an energy density functional of the form
\be
{\cal V}[{\bf Q}] =E({\bf Q}) -\frac {\tilde \kappa} 4 {\rm Tr}[{\bf Q}
{\bf D}{\bf Q} ] - \frac {{\tilde \kappa}'} 4 {\rm Tr}[ {\bf Q}^2{\bf D}{\bf Q} 
]
 +\ldots 
\label{FRANK}
\ee
where $D_{i,j}\equiv \partial_i\partial_j$ (neglecting total derivative terms).  
In the ordered phase, Eq.\ \ref{FRANK} leads to two
elastic moduli (Frank constants), but since they are interchanged by a 
90$^\circ$ rotation, the
difference between the two is  proportional to $Q$, and so
is small so long as $Q$ is small.  The dynamics of the
collective modes, as well as explicit expressions for the various coefficients 
which enter the
theory, must be derived from microscopic considerations.

\section{The Model}
\label{model}

We take as our model 
\ba
H
&=&\int d\vec{r} {\Psi}^{\dagger}(\vec{r}) \epsilon(\vec{\nabla} )\Psi(\vec{r}) 
\\
& & + \frac{1}{4}\int d\vec{r} \int d\vec{r}' 
F_2(\vec{r}-\vec{r}'){\rm Tr}[\hat{\bf Q}(\vec{r})\hat{\bf 
Q}(\vec{r}')].\nonumber
\ea
Here $\epsilon(\vec k)$ is the single-particle energy and we have ignored all 
density-density
interactions other than the essential ones, for present purposes, involving the 
quadrapolar
density.\cite{caveat} 
The single particle energy can be linearized about the Fermi surface, but for
later convenience we keep one further term in the expansion\cite{band},
$\epsilon(\vec k)ƒ=v_{F}q[1+a(\frac{q}{k_{F}ƒ})^{2}]$ with $q\equiv |\vec k|-
k_F$.  To be
explicit, we take the interaction to be the Fourier
transform of a simple Lorentzian, $F_2(\vec r) =(2\pi)^{-2}\int d\vec k e^{i\vec 
q\cdot \vec
r}F_2/[1 +\kappa F_2 q^2]$, where
$F_2$ is the appropriate Landau parameter.  Our results are not qualitatively
sensitive to any of these details. 

Landau parameters in a strongly correlated fluid are notoriously
difficult to deduce from microscopic considerations, but can be large.
In He$_3$, for instance, $N_F F_0$ is found to vary\cite{negel} from 10
to 80 as a function of pressure between 0 and 27 atmospheres.   At the
simplest level, one might hope to express these parameters in terms of
the Fourier transform of an effective density-denisty interaction as
\be
F_n= V(|\vec q|=0) \delta_{n,0} - \int_0^{2\pi} \frac {d\theta} {2\pi}
\cos(n\theta) V[2k_F\sin(\theta/2)].
\ee
Clearly, it is possible to obtain a large negative $F_2$ and positive
(or small) $F_0$ and $F_1$ from this expression, even if $V$ is
positive, especially if $V$ is peaked at a
 momentum transfer of order $2k_F$.  Such structure will occur
in any fluid with a large degree of local crystallinity.

To analyze the collective properties of this system,
 we introduce a Hubbard-Stratanovich \cite{book,kineq} field ${\bf n}$ to 
decouple the
four-fermion interaction, and then integrate out the fermions formally
to obtain the effective action, $S_{\rm eff}[{\bf n}]$.
Although ${\bf n}$ is also
a traceless symmetric tensor, it is convenient to introduce a 
vectorial notation in which
${\bf n}=n_{1}{\bf \sigma_{z}}+n_{-1}{\bf \sigma_{x}}$ where ${\bf 
\sigma}_{\alpha}$ are the Paul matrices.
While $S_{\rm eff}$ is very complicated, we can readily find the saddle-point 
solutions,
$\bar{\bf n}$, which are extrema of 
$S_{\rm eff}$, and can then  obtain explicit expressions for  
$S_{\rm eff}$ in powers of $\delta {\bf n}\equiv {\bf n}-\bar{\bf n}$.
In the limit $\kappa \rightarrow 
\infty$, the saddle-point (Fermi liquid) approximation becomes exact, 
and more generally this approach can be viewed as an expansion based
on the small parameter, $1/\kappa$.

~From now on we will work with the Hubbard-Stratonovich field $\bf n$ rather 
than with the order parameter field $\bf Q$. They are related by a Legendre 
transform,
and symmetry constrains the form of the effective actions in similar 
ways.\cite{book}
In particular, the effective action for the $\bf n$ field has elastic terms 
similar 
to those of Eq.\ \ref{FRANK} with new ealstic moduli $\kappa$ and $\kappa '$ 
which are 
proportional to $\tilde \kappa$ and ${\tilde \kappa}'$. 

The saddle point equations are obtained by minimizing an expression of the form 
of Eq.
\ref{GS2D}, with $\bar{\bf n}$ replacing ${\bf Q}$ and 
$A=\frac{1}{2N_F}+F_2$ where $N_F$ is the density of states at the Fermi 
surface.
Clearly, the isotropic phase is stable\cite{pom} so long as $2N_FF_2>-1$, while 
the
nematic phase occurs where this inequality is violated.  The quartic term $B$ is
determined by couplings that are formally irrelevant (in the renormalization 
group sense) at
the isotropic Fermi liquid fixed point;  for the explicit Hamiltonian considered 
above,
$B=\frac{3aN_F|F_2|^3}{8 E^2_F}$ where $E_F\equiv v_Fk_F$ is the Fermi energy.

 The effective action which governs the fluctuations 
 about the saddle point can be
computed to quadratic order in $\delta{\bf n}$,
\ba
&&\lefteqn{S_{\rm eff}[{\bf n}] = S_{\rm eff}[\bar {\bf n}]} \nonumber\\
&+& \frac{1}{2N_{F}}\sum_{a,b=\pm 1}\int
\frac{d\omega}{2\pi}
\frac{ d^{2}q}{(2 \pi)^{2}} \delta n_{a}^{*} {\Cal 
L}_{a,b}(\vec{q},\omega;\bar {\bf n}) \delta n_{b} + \ldots ,
\label{Seff-n}
\ea
where ${\cal L}_{a,b} $ is the inverse propagator of the collective modes, and 
the collective mode dispersion relation, $\omega_q$, is determined from the 
solution of the implicit equation
\be
{\rm det} [{\cal L}_{a,b}(\vec q,\omega)]=0.
\ee
It is convenient to express ${\Cal L}_{a,b}$ 
as the sum of  a static and dynamical piece: 
${\Cal L}_{a,b}(\vec q,\omega)\equiv {\Cal L}_{a,b}(\vec q,0)+
\tilde{{\Cal M}}_{a,b}(\vec q,\omega)$. 
 Symmetry
strongly  constrains the transverse component of ${\Cal L}_{a,b}(\vec q,0)
\sim q^2$ in 
the ordered phase. In other words,
being essentially the  static/classical deformation energy, its form is 
determined by the 
expression in Eq. \ref{FRANK}. However,
as we will see, symmetry considerations do not fully determine the dynamical 
piece. 

\section{The isotropic phase}  
\label{isotropic}

We warm up by analyzing the fluctuation
spectrum in the isotropic phase for $2N_{F}F_{2}> -1$.
By explicitly computing the fermionic response functions, with
$ \delta\equiv-\frac{1}{2}-\frac{1}{N_{F}F_2}$ and $s\equiv 
\omega/v_{F}q$, we find
\ba
{\cal L}_{a,b}(\vec q,\omega)&=&\delta_{a,b} (\kappa q^2 
+\delta) +\tilde{{\cal M}}_{a,b} (\vec{q},\omega) \\
\tilde{{\cal M}}_{a,b}(\vec q,\omega)&=&\frac{s}{2} \int_0^{2\pi} \frac{d 
\theta}{2 
\pi}\frac{{\cal P}_{a,b}(\theta)}{s-\cos (\theta-\phi)}, 
\label{M}  \\ 
{\cal P}_{a,b}(\theta)&=&
\left(
\begin{array}{cc} 
2\cos^{2}2\theta   &   \sin 4\theta\\  
 \sin4\theta       & 2 \sin^{2}2\theta
\end{array}
\right)
\label{LMF}
\ea
where $\phi$ is the polar angle subtended by
$\vec{q}$.
Without loss of generality, we can set 
$\phi=0$, in which case the two components of the pseudo-vector
correspond to the longitudinal ($n_{1}$) and transverse ($n_{-1}$) 
polarizations of the quadrupolar wave, and ${\cal L}_{a,b}$ is diagonal:
\ba
\tilde{{\Cal M}}_{a,b} (\vec q,\omega)&=&\delta_{a,b} {\Cal M}_{b} (s),\ \
{\rm with } \ \ s\equiv \omega/v_F q
\\ 
{\Cal M}_{\pm 1}(s)&=&\frac{s}{2\sqrt{s^2-1}}(1\pm(-s+\sqrt{s^2-1})^4)
\label{M1}
\ea
For  $F_2 > 0$ there exist
sound-like propagating  modes with $\omega_q > v_{F}q$ ( $s>1$), 
{\it i.\ e.\/} the 
quadrupolar analogs of zero-sound (one for each polarization). 
They are undamped because they lie 
outside of the particle-hole continuum. Closer to the nematic phase 
boundary, the evolution of the quadrupolar oscillations of the isotropic 
Fermi liquid deviates markedly from that of simple zero-sound. 
The dynamics clearly distinguishes the two polarizations and there are 
underdamped modes even when $F_2<0$.

\section{The quantum critical regime} 
\label{quantum-critical}

The difference in the dynamics of the 
two polarizations becomes 
more pronounced as the quantum critical
point is approached. As $\delta\rightarrow 0^{+}$ and for  $\omega \ll v_F q$
\ba
{\cal L}_{++} (\omega,q)&=&\kappa q^2+\delta - i\frac{\omega}{v_F 
q}+\cdots;\\
{\cal L}_{--} (\omega,q)&=&\kappa q^2+ \delta - (\frac{\omega}{v_F 
q})^2-2i(\frac{\omega}{v_F 
q})^3+\cdots.
\ea
The 
transverse mode 
becomes more and more weakly damped with $\omega_- \approx i\sqrt{\delta}v_F q 
+v_f\sqrt{\kappa}q^2$ while the longitudinal mode remains overdamped, 
$\omega_+ \sim iq^3$.   This behavior should be compared and contrasted with the
behavior of the paramagnon collective mode near the Fermi liquid to 
ferromagnetic 
Fermi liquid  transition where, in the disordered phase and at the 
critical point, all three polarizations are identical, and indeed have
dispersions similar to that of the  longitudinal (``++'') quadrupolar mode. 
The difference arises from the fact that, unlike in the nematic, in the ferromagnet
(in the absence of spin-orbit coupling), the broken symmetry is
unrelated to any spatial symmetries.
Nevertheless, as with the ferromagnet, a straightforward scaling
analysis of the effective action implies a
dynamical critical exponent $z=3$ at the quantum critical point.
Remarkably, at the critical point, the transverse mode has higher
characteristic energy ($\omega_{-}\sim q^{2}\sim \omega_{+}^{2/3}$),
than the overdamped critical mode;  it plays no role in
the critical theory, and indeed the $\omega^{2}/q^{2}$ term in the inverse 
propagator which
makes it  dynamical is  irrelevant for 
$z=3$ scaling, 
$(\frac{\omega}{v_F q})^2 \sim  q^4$.  Moreover, $z=3$ also implies 
that 
interaction terms of order ${\bf n}^4$ and higher in $S_{\rm eff}$ are 
irrelevant at the quantum critical point. 
Thus, according to the 
standard lore\cite{hertz,millis}, the critical behavior is fully 
captured by the gaussian theory.

\section{The nematic phase} 
\label{nematic}

In the  ordered phase, but close to the quantum critical point, 
the order parameter is small.  Here we can, to good approximation,
ignore the dependence of the $\vec q$ and $\omega$ dependent terms in 
the effective action on the  ordered moment\cite{caveat2}
${\cal L}_{a,b}(\vec q,\omega;\bar {\bf n} )=
{\cal L}_{a,b}(\vec 0,0;\bar {\bf n} )+ {\cal L}_{a,b}(\vec q,\omega;{\bf 
0} ) +\ldots.$
The  higher order terms are, among other things,  
responsible for the  difference between the two Frank
constants 
in the elastic energy of the nematic;  
as they complicate the normal mode analysis, and 
make little important qualitative difference, we will defer 
considering them until later.
Without loss of generality,
we can choose the principal axis of the nematic state to lie along 
$\hat x$, so that $\bar {\bf n} =\bar n {\bf \sigma_{z}}$ 
(where ${\bf \sigma_{z}}$ is the Pauli matrix).
In this case, fluctuations of the amplitude of the order parameter
are associated with the $+1$ component of  $\delta 
{\bf n}$, and $-1$ are the phase ({\it i.e.} orientational) 
deformations.  
Inside the  nematic phase, as for phonons in a crystal,
the longitudinal
and tranverse modes are mixed unless $\vec q$ lies along a symmetry 
axis.  

Because we have neglected the difference between Frank constants, 
the purely elastic energy is a sum of independent contributions from 
the phase and amplitude modes,
\ba
{\cal L}_{++}(\vec q,0;\bar {\bf n})  & = &
2|\delta| +\kappa q^2   \nonumber\\
{ \cal L}_{--} (\vec q,0;\bar {\bf n})
& = & \kappa q^2,
\ea
and ${\cal L}_{-+} (\vec q,0;{\bf 0})=0$. However, even at this level of 
accuracy, 
the dynamics unavoidably mixes the phase and amplitude modes.  
Explicitly, the dynamical matrix is a function of the scaled variable
$s=\omega/v_{F}q$ {\it and} the angle, $\phi$, between the principal 
axis of the nematic order and the wave-vector $\vec q$ 
\be
{\cal M}(s,\phi)=
\frac{s}{2}
\left(
\begin{array}{cc}
B(s)+A(s)\cos 4\phi & A(s)\sin 4\phi \\
A^{*}(s)\sin 4\phi & B(s)-A(s)\cos 4\phi
\end{array}
\right),
\ee
with $B(s)=1/\sqrt{s^2-1}$ and $A(s)=B(s)(\sqrt{s^2-1}-s)^4$.  ($A^*(s)$-the complex 
conjugate of $A(s)$.)
Only if $\vec q$ is along 
a symmetry direction, $\phi=0,\pm \pi/4,\pm \pi/2,$ etc., is ${\cal M}$
diagonal.  That the longitudinal and transverse modes are
mixed for all other propagation directions,
even in the absence of terms which 
depend explicitly on $\bar {\bf n}$ is, at first sight, curious, and 
is a direct consequence of  the fact that the nematic order parameter breaks
spatial rotational symmetry, not an internal symmetry (such as spin).  

As in the isotropic phase, we diagonalize ${\cal L}_{a,b}$ to 
obtain the collective mode spectrum.   
An excitation is called ``transverse'' if for $\vec{q}\rightarrow0$ it 
is polarized perpendicular to the principal axis of the nematic and 
``longitudinal'' otherwise; the corresponding eigenvalues of the 
inverse propagor are
${\cal L}_{\perp}$ and ${\cal L}_{\parallel}$.  
  
For general $\phi$ and as $s\rightarrow0$, the inverse propagator for the 
transverse (Goldstone) mode is
\be
{\cal L}_{\perp} = \kappa q^2 -
\frac{i s}{2} \sin^2 2\phi - s^2 
\left(
\cos 4\phi+
\frac{\sin^2 4\phi}{128 |\delta|}
\right)
+ {\cal O}(s^3)
\label{diff}
\ee
This result is remarkable and, to our knowledge, unprecedented: 
Landau damping dominates the dynamics of the Goldstone boson, making 
it overdamped (the last term is irrelevant) over most of  phase space. 
Formally, Eq. \ref{diff} is reminiscent of those 
studied in conjunction with quantum criticality in metals\cite{hertz,millis}, 
or when a 
transverse gauge field is coupled to a Fermi liquid 
\cite{HNP,ganwong}.
In all three cases Landau 
damping results in a soft boson with dissipative dynamics.  
Naturally, the origin of the softness is different in each case;
while the  softness of the quantum critical propagator is the result 
of fine 
tuning to criticality, the softness of the gauge and the 
director-wave 
propagators stems, respectively, from gauge invariance and broken 
rotational
symmetry. Nevertheless, here, as in the other two problems (and in two dimensions), Eq. \ref{diff} implies a specific heat vanishing as $T^{2/3}$ at low temperatures.

The angular dependent damping term in Eq.\ \ref{diff} is 
native to the nematic state; along symmetry directions, 
when it vanishes, a propagating ({\it i.e.}undamped as $q\rightarrow0$) mode results.
For $\phi=0$ and $\frac{\pi}{2}$
, and for $\omega\ll v_Fq$
\ba
{\cal L}_\perp^{0,\frac{\pi}{2}}= \kappa
q^{2}-s^2 -i 
s^{3}+{\cal O}(s^4),
\label{wq2}
\ea
implying a dispersion according to $\omega_q=v_F\sqrt{\kappa}q^2$.  An additional propagating mode with a soundlike spectrum $\omega_q=\frac{v_Fq}{\sqrt{2}}$ exists for $\phi=\frac{\pi}{4}$.
These longlived modes are somewhat peculiar: they are already present at 
the quantum critical point (with precursors even in the isotropic phase!), 
where indeed they propagate in all 
directions,
although 
they appear as higher energy excitations which do not directly enter 
the critical phenomena.  

These collective phenomena can be summarized pictorially:
\begin{figure}
\begin{center}
\leavevmode
\epsfxsize=3in
\epsfysize=2.5in
\epsfbox{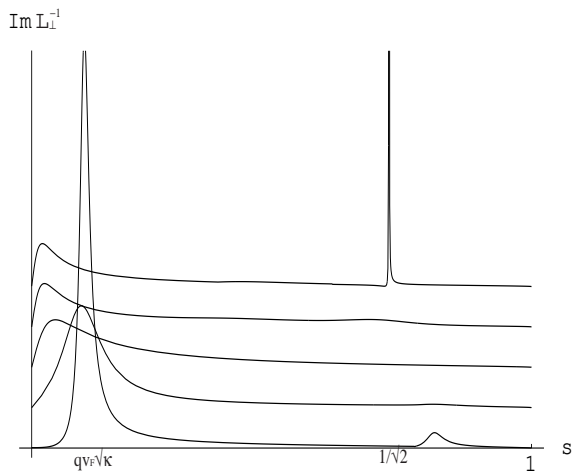}
\end{center}
\caption
{The spectral function, $Im \chi = Im {\cal L}^{-1}_\perp$, of the nematic Goldstone modes as a
function of $s=\omega/v_F q$ (at fixed small $q$) for different angles of 
propagation from (top to bottom) $\phi=\pi/4$, $\phi=3\pi/16$, $\phi=\pi/8$,  
$\phi=\pi/16$ and $\phi=0$, respectively. The curves are plotted on the same scale, with 
vertical offset, for $\delta=0.1$, $\kappa=1$, $v_F=1$ and $q=0.15$.
At $\phi=0$ there is a sharp propagating mode (see eq. \ref{wq2}). 
Otherwise the spectral function is linear as $\omega\rightarrow0$, 
signifying a diffusive peak ($\frac{Im \chi (\omega)}{\omega}$ is 
peaked at $\omega=0$). At $\phi=\pi/4$ the spectrum contains an additional long 
lived 
soundlike mode.
}
\label{fig:fig1}
\end{figure}

Finally the inverse propagator of the amplitude mode is
\be
{\cal L}_{\parallel}= 2|\delta| - \frac{is}{2}\cos^2 2\phi + {\cal O}(s^2). 
\label{long}
\ee

The apparent four fold symmetry in these expressions
is, in part, inherited from the 
special symmetry of the two dimensional director: $\bar {\bf n}$ is
odd under
rotation by $90^\circ$ so 
quantities even in $\bar {\bf n}$ will be {\it four} fold symmetric, even inside 
the 
nematic state.  However, the precise four-fold symmetry is an artifact 
of our neglecting the $\bar {\bf n}$ dependence of the $\vec q$ and 
$\omega$ dependent terms in ${\cal L}$.  For instance, $\kappa$ should rightly 
be replaced by the two distinct Frank constants, 
$\kappa_{\perp}$ and $\kappa_{\parallel}$ in Eq. (\ref{diff}), 
respectively, where from Eq. \ref{FRANK}, $\kappa_{\perp}-\kappa_{\parallel}\sim 
\kappa'
|\bar {\bf n}| + \ldots$.  At the level of the gaussian theory constructed thus 
far 
the corrections can be analized perturbatively (in $\bar{\bf n}$), leading neither to 
qualitative modifications of modes' dispersions nor to spoiling of the fourfold symmetry (e.g. in the damping term in Eq. \ref{diff}).
More generally, because the nematic order does not gap any section of 
the Fermi surface, but only distorts its shape, the dynamical 
consequences of the fermionic particle-hole continuum are preserved, 
even far from the critical point. In other words, the characteristic scale for 
the frequency dependence is still $v_F q$ ({\it {i.e.}} vanishing at long 
wavelengths) in the broken symmetry phase,
regardless of the magnitude of the Fermi surface distortion.

Clearly, interactions among the collective modes can be treated perturbatively 
in $1/\kappa$. What the effects of these interactions are away from the critical 
point is presently unknown. 

It is important to recall that any symmetric traceless tensor can 
serve as the order parameter for the nematic state.  While we have 
chosen the deformation of the Fermi surface, alternate choices
include the inverse mass tensor which defines the Drude weight 
of the optical conductivity,  
and the resistivity tensor, itself.
They can be interrelated explicitly in our 
case (the corrections to this first order result are cubic)
\be
\frac{\rho_{xx}-\rho_{yy}}{\rho_{xx}+\rho_{yy}}=\frac{1}{2}\frac{m_y-
m_x}{m_y+m_x}=\frac{n_{11}}{E_F}  + {\cal O}(|{\bar n}|^3)
\label{mass-anisotropy}
\ee
where $m_x$ and $m_y$ are the effective masses of the quasiparticles in the 
nematic state, as determined by the spontaneous anisotropy of the Fermi surface.

\section{Single particle properties}
\label{single}

We will now consider the effect of the collective modes on the single-particle 
self energy.
To one loop order, the imaginary part of the self-energy (the scattering rate) 
is given by\\
\ba
\Sigma''(\epsilon,\vec{k}) &=&\frac{\pi}{\sqrt{3}}\frac{(\kappa k^2_F)^{1/3}}{\kappa N_F}
\left | \frac {k_{x}k_{y}} {k_F^2} \right|^{4/3}
\left |\frac{\epsilon}{2 v_F k_F}\right|^{2/3} +
\ldots
\label{S}
\ea
where $\vec k$ lies on the Fermi surface and $\ldots$ signifies subleading terms
in powers of $\epsilon$.
The strong angular dependence of $\Sigma^{''}$ is  a startling consequence of 
the symmetry
of the nematic. Along the symmetry directions, $\vec{k}=(k_x,0)$ and $(0,k_y)$, 
the scattering rate (at the Fermi surface) has a different energy dependence 
indicative of a long lived quasiparticle\cite{caveat4}: 
\ba
\Sigma''(\epsilon)&=&
\frac{\pi}{3 N_{F} \kappa} \frac{1}{(\kappa k^2_F)^{1/4}}\left | \frac{\epsilon}{v_F k_F} \right |^{3/2}+\ldots
\ea 

Although perturbative, our results unambiguously signal the breakdown of Fermi
liquid theory, {\it i.\ e.\/} the spectral function
no longer has a quasiparticle pole over most of the Fermi surface\cite{caveat4}.  
For fixed $\epsilon$, the perturbation theory is
arbitrarily accurate for large enough $\kappa$, but for fixed $\kappa$, 
perturbation
theory may well break down at small $\epsilon$.  We have not determined yet 
whether this indicates a phase transition to a new, ordered state, {\it e.g.} a
superconducting state, or the occurrence of a genuine, two dimensional non-Fermi 
liquid phase.
With the exeption of the discrete set of symmetry related points where the 
quasiparticle survives collective damping, the frequency dependence of our  
result is
identical to the comparable perturbative result for fermions coupled to 
(Landau damped) gauge or quantum critical
fields\cite{hertz,millis,HNP,ganwong}.
However, it is important to emphasize that formal resemblance  among these three 
problems need not persist beyond the lowest order in perturbation theory.

\section{Extensions and speculations} 
\label{extensions}

Thus far we have considered a rotationally invariant system. 
Since the electron fluid is typically realized in a solid state context, it is
important to consider the effects of explicit  rotational symmetry 
breaking by the underlying lattice.
These are relevant operators in the nematic phase, and their main 
effect is to gap out the Goldstone bosons.
 In the presence of such explicit symmetry breaking, 
 the system behaves, at sufficiently low energy, like an
anisotropic metal with Fermi liquid properties. However, an analysis 
of their effects on the quasiparticle
self-energies indicates a crossover from non-Fermi liquid  to Fermi liquid
 behavior below an energy scale
$\epsilon^{*}\approx \frac{v_F
\lambda^{3/2}}{\sqrt{\kappa}}$ where $\lambda$ is a dimensionless measure of the
lattice induced anisotropy on the bare quasiparticle energy.  
In many cases,  $\lambda$ is small so $\epsilon^*$ is very small. 

There are many obvious and interesting generalizations of these ideas.  With 
very little
difference, we can consider the theory of the metallic hexatic, which is 
similarly
non-Fermi liquid.  Such a state would be triggered by a large, negative $F_6$.  
While at
first that might sound artificial, if we view the quantum hexactic as a melted 
Wigner crystal,
it is a natural state to consider.  This may have relevance to the long-standing 
problem
of apparently metallic states in two dimensions.  In the presence of electron 
spin,
a host of new and interesting states become possible which intertwine spin and 
spatial ordering.
For example, one could imagine a state triggered by a large and repulsive 
$F_2^a$
({\it i.e.} in the triplet channel) in which the nematic order parameters of
the spin-up and spin-down electrons are rotated relative to each other.
Finally, deeper inside the nematic phase the Fermi surface gets increasingly distorted, 
leading to nearly nested segments; it is thus natural to consider a 
further instability, triggered by ``backscattering'', to a stripe ordered 
electronic smectic phase.

The three dimensional nematic can be analyzed similarly.
Here, the director is a
$3 \times 3$ rank 2 symmetric traceless tensor.
On symmetry grounds one anticipates that, since
a cubic invariant is allowed in the free energy, the isotropic to nematic 
transition
will generally be  first order. We have repeated the calculation of the ground 
state energy
(Eq.  \ref{GS2D}) in three dimensions. 
In contrast to two dimensions, we indeed find a cubic term which usually 
\cite{deGennes} 
favors a uniaxial 
nematic over a 
biaxial one and its sign determines whether the nematic Fermi surface
is oblate or prolate. The Goldstone mode is overdamped as in 2D; its 
contribution to the 
single particle scattering rate is of the ``marginal Fermi liquid'' 
form\cite{varma}, 
$\Sigma'' (\epsilon)\propto |\epsilon|$, but strongly angle dependent.

The considerations presented in this paper may serve as a starting point
of a microscopic theory of a quantum Hall nematic fluid.
In
particular, it was proposed in ref.\ \ref{ref:FK} and
\ref{ref:FKMN} that the observed large anisotropy in Landau levels $N \geq 2$ 
could be
explained by a nematic fluid phase. The results that we have derived in this 
paper 
indicate
that this is a very concrete possibility. Moreover, the relation between
anisotropic transport and a nematic order parameter  proposed on symmetry 
grounds
 in ref. \ref{ref:FKMN} 
follows clearly from the spontaneous effective mass anisotropy in the nematic 
state 
(discussed
above) by means of simple Boltzmann transport arguments. 
A candidate grounstate wave function for a quantum Hall nematic at filling 
factor 
$\nu=1/2$, can be constructed
in the spirit of the Jain wavefunction for the Composite Fermi 
Sea\cite{jain,rezayi}, as
\be
\Psi_{CNFL}(z_i)={\cal P}\prod_{i>j} (z_i-z_j)^2 e^{-\sum_j 
\frac{|z_j|^2}{2}}\Psi_{NFL},
\ee
where $\Psi_{NFL}$ denotes the ground state wavefunction of the 2D nematic we 
have considered here, $z_i$ are the complex coordinates of the electrons (with 
the 
magnetic
length set to 1) and ${\cal P}$ is the Landau level projection operator. 
However, 
for a subtle 
gapless state like this it is not clear that simple wave functions are able to 
fully 
capture the physics.

Finally, the picture of the nematic phase that emerges from our results is
strikingly reminiscent of the behavior of the so-called ``normal phase" 
of high temperature superconductors. In particular the fermion spectral
function that we find in the nematic phase has a behavior that is
qualitatively similar to the electron spectral function measured
by ARPES in BSCCO\cite{shen,campuzano,valla}. In fact, L.\ B.\ Ioffe and A.\ J.\ 
Millis\cite{ioffe}
have proposed a phenomenological form for the quasiparticle scattering rate with
the angular dependence $|k_x- k_y|$ and a $|\epsilon|$ marginal Fermi liquid 
energy dependence.
It is tempting to speculate that a possible origin of this sort of behavior may 
be a 
nematic Fermi fluid state setting in at temperatures close to the pseudogap 
$T^*$. In fact, in Ref. \ref{ref:KEF},
 two of us suggested that as a strongly correlated 
electron system 
is cooled down from a high temperature isotropic fluid phase, the first
stage of charge ordering should be precisely nematic phase\cite{caveat3}. As 
stressed
in ref.\ \ref{ref:KEF}, this is a Kosterlitz-Thouless thermodynamic phase 
transition,
which is reduced to an Ising transition in by fourfold lattice anisotropy or 
rounded in an orthorhombic crystal. Incidentally, spontaneous breaking of rotational symmetry in the two-dimensional Hubbard model has also been studied numerically\cite{halb}. Recent light scattering
experiments  by M.\ Rubhausen and coworkers\cite{lance}
have found evidence of a rounded nematic transition at the charge-ordering 
temperature
of the manganite $Bi_{1-x} Ca_x MnO_3$.  

\section*{Acknowledgements}

We wish to thank J.~Zaanen,
C.~Nayak, S.~Girvin, S.~Chakravarty, J.~C.~Campuzano and K.~Bedell for useful discussions.
We were participants in the High $T_c$ Program at ITP-UCSB.
This work was supported in part by NSF grant numbers DMR98-08685
at UCLA (SAK and VO), NSF DMR98-17941 at UIUC (EF), and PHY99-07949 at
ITP-UCSB.

\end{multicols} 

\begin{references}
\bibitem{caveat}
We have verified that the interactions in other channels, which are always
present, in particular $F_0$
and $F_1$, do not change in any essential way our conclusions.
\bibitem{band}
This is the first correction that stabilizes the nematic state.
\label{band}
\bibitem{kineq} 
Of course, identical results are obtained using the more traditional, 
kinetic equation approach.
\bibitem{caveat3}
A potentially serious flaw with this idea is that the most natural orientation 
of the nematic order would have symmetry axes at $45^\circ$ to the direction 
implied by ARPES experiments in BSCCO; a potential resolution of this problem 
would be to consider the possibility of ``tetratic'' order.
\end{references}
\end{document}